\documentclass[11pt]{article}

\usepackage{graphics}
\usepackage{pdfpages}
\usepackage[backend=biber,style=alphabetic,]{biblatex}
\begin{document}

\title{Varying Annotations in the Steps of the Visual Analysis}

\author{\parbox{\textwidth}{\centering C. Schmidt$^{1}$ 
        and P. Rosenthal$^{1}$ 
        and H. Schumann$^{1}$ 
        }
        \\
    {\parbox{\textwidth}{\centering $^1$University of Rostock, Germany\\
       }
    }   
}

\maketitle


\begin{abstract}
Annotations in Visual Analytics (VA) have become a common means to support the analysis by integrating additional information into the VA system. 
That additional information often depends on the current process step in the visual analysis.
For example, the data preprocessing step has data structuring operations while the data exploration step focuses on user interaction and input.
Describing suitable annotations to meet the goals of the different steps is challenging.
To tackle this issue, we identify individual annotations for each step and outline their gathering and design properties for the visual analysis of heterogeneous clinical data.
We integrate our annotation design into a visual analysis tool to show its applicability to data from the ophthalmic domain.
In interviews and application sessions with experts we asses its usefulness for the analysis of patients with different medications.
\end{abstract}  


\section{Introduction}
\label{010_introduction}

    Data preprocessing, data cleansing, and data exploration are common steps in visual analytics processes \cite{Sacha2017}, \cite{Schmidt2019}, \cite{Gschwandtner2012}. 
    Each of these steps has its own challenges. 
    For data preprocessing often several data sources need to be considered in order to fuse the data to be analyzed.
    This data fusion may lead to redundant and possibly contradictory data values. 
    For data cleansing, the detected data discrepancies and incompletenesses need to be solved in order to generate a consistent dataset. 
    For data exploration, interesting data characteristics need to be marked, commented and/or extracted in order to generate new insights and possibly new knowledge.

    While annotations have proven useful to support the overall visual analysis \cite{Zhao2017}, specific support for the different steps in the analysis is ongoing research.
    This regards, for example, how annotations can help to (i) mark and communicate data redundancies and discrepancies during data preprocessing, (ii) support users on data cleansing decisions, and (iii) externalize and/or comment findings during exploration.
    With our approach we tackle these questions by designing a concept on annotation usage in the different steps. 
    In this concept we describe automatically generated annotations on data value redundancy information and discrepancy resolution . 
    During data cleansing we integrate annotations that allow users to explain decisions on solved discrepancies and to track changes made. 
    With annotations for the exploration step, we support the externalization of findings, the recording of user knowledge as well as the integration of further user comments, e.g. comments that judge previous annotations.

    With these annotations at hand, we extend an existing visual analytics tool (from \cite{Schmidt2019}), allowing the gathering, processing, and communication of additional information that assists users during their visual analytics process.
    We apply the extended tool on a use case involving heterogeneous, contradictory, and incomplete data from the medical domain. 
    In doing so, domain experts are able to process and analyze data for several thousand patients. 
    This allows to draw medical conclusions by analyzing a single consistent data set instead of analyzing multi-center clinical data with multi-center data biases.

\section{Related Work}
\label{sec:0200_related_work}
    As Lipford et al.~\cite{Lipford2010} and Mahyar et al.~\cite{Mahyar2012} point out, the use of annotations can be a critical step in visual analytics.
    In literature there are many ways to show annotations for different purposes.
    This concerns for example the introduction of general classifications, as done by, e.g., van Hulst et al. \cite{Vanhulst2018}, Saur\'i et al \cite{Sauri2017}, or Schmidt et al \cite{Schmidt2018}.

    More specific applications depict annotations in the exploration step of the analysis.
    This concerns, for example, Willett et al. \cite{Willett2011} and Mahyar et al. \cite{Mahyar2014}, who use annotations to enable free text comments and communication between different users.
    The work of Elias and Bezerianos \cite{Elias2012} supports data and context aware annotations. They focus on multi-target annotations, annotation transparency across charts and data dimension levels, as well as annotation lifetime and validity.
    Zhao et al. \cite{Zhao2017} concentrate on annotation examination by introducing annotation graphs to communicate insights between analysts.
    Other examples for annotations in data exploration are Heer et al. \cite{Heer2007} and Zhao et al. \cite{Zhao2018}, who concentrate on annotations for asynchronous collaborative analysis, Shabana and Wilson \cite{Shabana2015} who present a novel scalable method for automatic discovery, annotation and interactive visualization of prominent segments in mobile subscriber datasets, and Groth and Streefkerk \cite{Groth2006} who present a model for recording the history of user explorations in visualization environments, augmented with the capability for users to annotate their explorations.

    Also, the data preprocessing step (Schmidt et al. \cite{Schmidt2019}) and data cleansing step (McCurdy et al. \cite{McCurdy2018}) are supported by annotations.
    Despite the fact, that literature shows various work on annotations supporting steps in the analysis, work on several steps at once is still challenging.
    With our work we focus on covering several steps at once with differentiated annotations.
    
\section{Annotation Description for the Analysis Steps}
\label{030_process_driven_annotations}
    In this section we show, how specific annotations can address typical problems in different steps of a visual analysis process.
    This particularly concerns the data preprocessing, data cleansing, and data exploration steps.
    We chose these steps, as previous analyses in the field of heterogeneous real-world data have identified them as important (e.g. \cite{Gschwandtner2012}, \cite{Schmidt2019})

    \subsection{Data Preprocessing Annotations}
    \label{033_annotations_for_data_preprocessing}
        \begin{figure}[ht]
            \centering
            \setlength{\fboxsep}{0pt}
            \setlength{\fboxrule}{0.5pt}    
            \fbox{\includegraphics[width=0.7\linewidth]{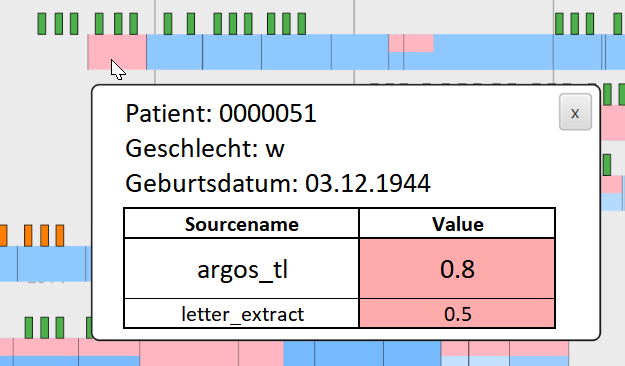}}
            \caption{\label{fig:anno_preprocessing}
                Visual encoding of annotations during data preprocessing. The information, if there are redundancies and/or discrepancies, is encoded by color in the data visualization. On demand, details of the annotation content is displayed.
            }
        \end{figure}

        The data preprocessing step has the goal, to collect and structure necessary information from all available data sources.
        These data may stem from different sources, as real-world data often are collected from various devices and scientific data often are redundantly recorded to decrease uncertainty.
        When these sources are merged, redundancies and discrepancies may appear.
        With data-preprocessing annotations, we automatically collect that additional information on source redundancy and discrepancy for each data value.
        These annotations are linked to the respective data-point and amended with the source name and time stamp.

        Additionally, data sources may differ in reliability.
        This is specifically the case, if a data source contains primary and secondary information.
        The primary information is the main information (e.g. data dimension) from a source with a high degree of reliability. 
        The secondary information has also been recorded within the data source, yet is not in the focus and may thus be less reliable.
        An example for clinical data may be from the clinical management system. 
        There the diagnoses may be primary information, as it is important for billing purposes, while the pathologies are also kept to inform the patient or relatives, but are not used by physicians and are therefore secondary information with less reliability.
        If this is the case, the experts need to define a hierarchy for each data dimension and data source, in order to automatically resolve discrepancies.
        That hierarchy information on which an automatic decision is based, is also saved by an annotation linked to the data value.
        As a result, we get annotated structured data, providing the information to the user, what data sources were considered, if redundancy and/or discrepancies apply and what rule lead to the choice of value.
        The annotations are made available to the user by a respective encoding at later stage in the analysis. 
        Figure \ref{fig:anno_preprocessing} shows an example with detail information on existing sources, including a value discrepancy (0.8 vs. 0.5).

    \subsection{Data Cleansing Annotations}
    \label{031_annotations_for_data_cleansing}

        \begin{figure}[ht]
            \centering
            \setlength{\fboxsep}{0pt}
            \setlength{\fboxrule}{0.5pt}    
            \fbox{\includegraphics[width=0.7\linewidth]{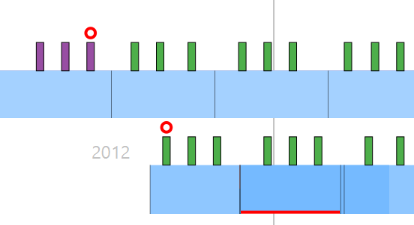}}
            \caption{\label{fig:anno_cleansing}
                Visual encoding of annotations created during editing. Edited data values are locally marked with varying glyphs, depending on the data range effected by the editing operation.}
        \end{figure}

        During data cleansing the user judges data values in regard to their plausibility and completeness.
        As data discrepancies are common for real-world data, this step often involves data editing in order to create a consistent data-set (see, e.g., \cite{Schmidt2019}).
        As the data editing functions can either be manual or automatic and data discrepancies can variously be resolved, annotations should inform the user when, how, and by whom this was done.
        We reach that by storing the rule set on which automatic data corrections have been performed.
        This information can then be shown to the user when necessary and with specific visualization techniques.
        An example is provided in Figure \ref{fig:anno_cleansing}.
        The annotated data is highlighted in the visual representation used for data cleansing (red circles and lines), and when clicking an annotated data value, a textbox with detail information is presented.
        In case of manual data corrections, a validation of the editing can be useful, especially in asynchronous collaborative environments as well as in discontinuous processing as described by Zhao et al. \cite{Zhao2018}.

    \subsection{Data Exploration Annotations}
    \label{032_annotations_for_data_exploration}

        \begin{figure}[ht]
            \centering
            \setlength{\fboxsep}{0pt}
            \setlength{\fboxrule}{0.5pt}    
            \fbox{\includegraphics[width=0.7\linewidth]{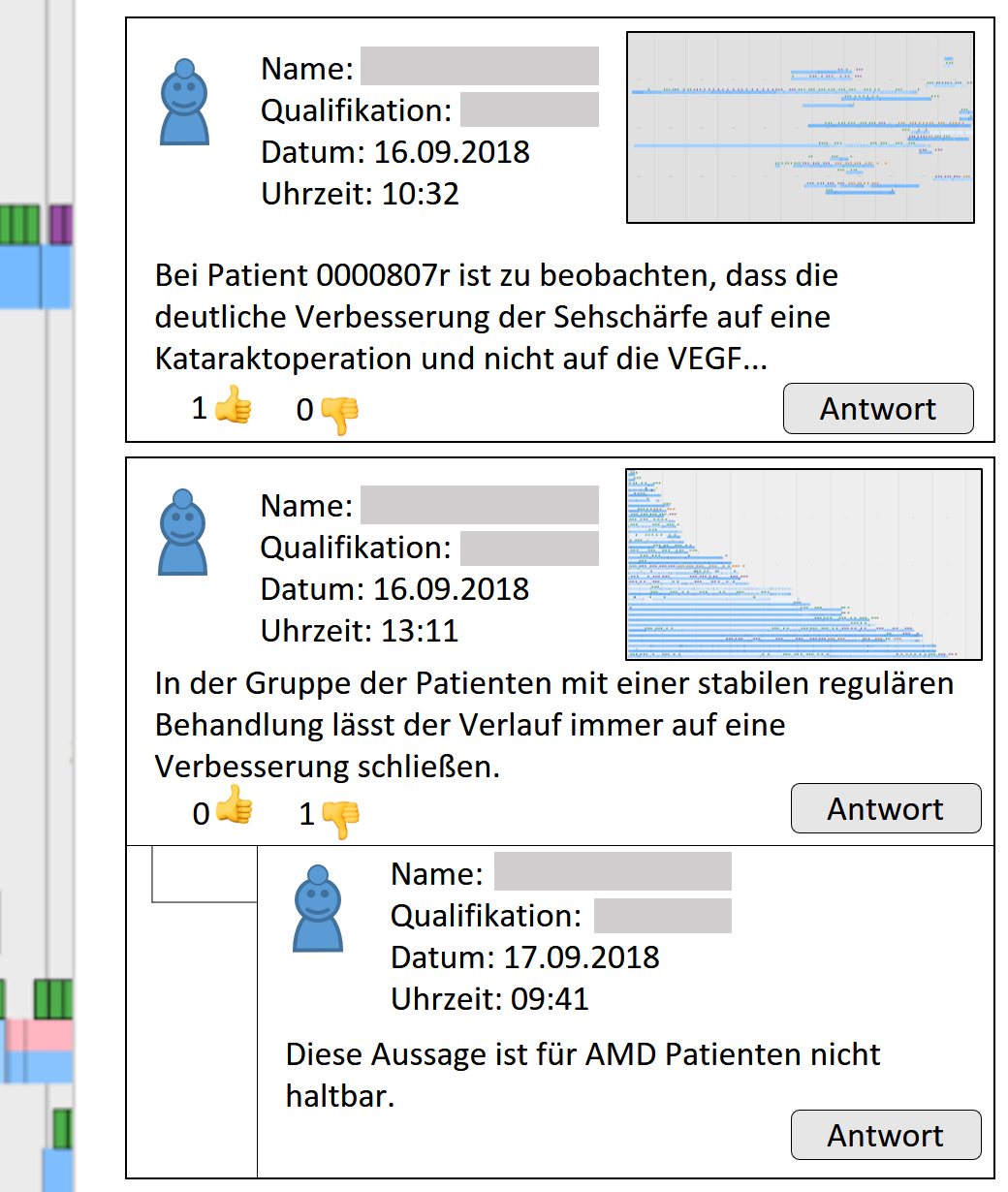}}
            \caption{\label{fig:anno_comment}
                The annotation area for comments and validation information. Previous Annotations can be commented or declared as validated/invalid.}
        \end{figure}

        With annotation support during data exploration, we allow users to take notes within the system. 
        This can be done for different purposes. First, identified findings can be marked and externalized via annotations. Second, existing externalizations, saved via annotations, can be commented or validated. Third, validated annotations can be separated by users from unvalidated or invalid annotations to support further analysis.
        To allow later interpretation of a marked and externalized finding, we provide (i) the verbalized finding from the expert, (ii) the visualization at the time of externalization via screenshot, and (iii) a reference to all data points visible at the time of externalization. 
        This will give users the ability to interpret the verbalization in combination with the visual representation at the time of externalization.
        Moreover the reference to the data, allows other users to check for updated data and/or annotations.
        At any time, recorded annotations are displayed in a separate view. 
        This way, we can provide commenting or validating functions for all annotations.

        Such functions are particularly important, if the exploration process is interrupted or performed by different users. Then, it can be necessary to judge or comment previously recorded annotations. 
        Commenting is applied, if the interpretation of early findings changes, for example since new information is available, or a user contributes new knowledge that leads to a different view of a finding. 
        In particular, we enable users of a certain qualification to make an assessment of existing externalised information (annotation of the annotation). 
        This way, annotations can be confirmed. 
        This facilitates a separation into unvalidated/valid/invalid annotations. 
        We present all annotations made in the exploration process (see Figure \ref{fig:anno_comment}) to support the interpretation of the findings and the generation of hypotheses. 

\section{Use Case}
\label{040_anno_in_model}
    Our use case is situated in the medical domain.
    We have heterogeneous clinical data stemming from several thousand patients.
    As the data is potentially redundant, has discrepancies and may be incomplete, we aim at extending an existing VA tool with additional annotation functionality that helps domain experts to address these issues.
    In this section, we show the architecture and design of our extended tool with a special focus on the integration of annotations, and depict the feedback from the domain experts.
    \subsection{VA Tool with annotation design}
        \begin{figure}[ht]
            \centering
            \setlength{\fboxsep}{0pt}
            \setlength{\fboxrule}{0.5pt}    
            \fbox{\includegraphics[width=0.7\linewidth]{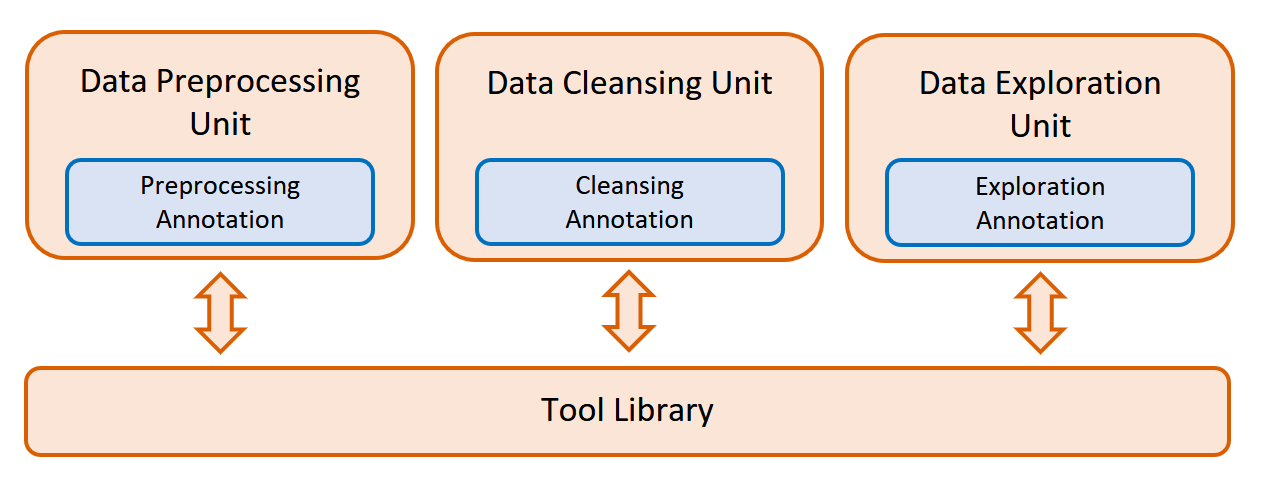}}
            \caption{\label{fig:anno_architecture}
                A schematic illustration of the VA tool architecture. For each step in the analysis, a dedicated unit is responsible for annotation management. This is supported by functions derived from the tool library.}
        \end{figure}

        In this subsection we briefly summarize the annotation functionality of our visual analysis tool. The tool itself has already been described by Schmidt et al. \cite{Schmidt2019}.

        Generally, all annotation management is integrated into the tool units as seen in Figure \ref{fig:anno_architecture}.
        Automatically generated annotations are added during data pre-processing, whereas manual annotations are recorded while data cleansing and exploration are performed.
        Automatically derived annotations are visible at the data preprocessing stage.
        On the other hand, data editing annotations are visible during data cleansing, and manual annotations on exploration are visible in that phase respectively.
        The annotations for each step have a specialized design, which is described below.

        Automatic annotations for data preprocessing are encoded via color by a separate layer (see Figure \ref{fig:anno_preprocessing}).
        The color for the annotation is distinct from the data color coding to emphasize annotated data.
        As preprocessing annotations encode the reliability of the data values: The colors of the layer occlude data values (i) completely, if there is a discrepancy, (ii) partly, if there is only one source, and (iii) not at all, if the data values are derived from several sources but no discrepancies occur.
        This has the positive effect that experts see the colored data regions with discrepancies at first sight, while they are not wrongly influenced by seeing a possibly incorrect data value.
        In addition, we show further details on demand, such as corresponding data sources, data values, or source priorities.

        During data cleansing annotations on editing operations are manually added by user interaction.
        We provide the annotation recording interface within the editing dialog.
        Once the recording has been performed, the annotated data are amended with additional glyphs in the visualization (see Figure \ref{fig:anno_cleansing}).
        To avoid confusion between data values and annotation indication, these glyphs are distinct in shape and color from the data encoding.
        The glyphs are depicted nearby the annotated data values, and by hovering them, the full annotation content will be presented via a separate view. 

        The annotation design during exploration is different from the previous steps.
        Here, annotations are often not directly related to single data values, but to particular data features.
        Moreover, we want to support the judgement of the displayed data and annotations.
        For that reason, we decided to separate the annotation design from the data visualization view.
        Instead, we provide global annotation functions with a separate view (see Figure \ref{fig:anno_comment}).
        Within that view all previously recorded annotations for exploration are shown.
        At this step, the experts focus on several points. 
        First, they want to judge who recorded an annotation when and with what qualification, so we show a short profile in the top left corner.
        Second, they need a visual impression on the visualization relevant for the annotation, which we provide by a thumbnail in the upper right corner, that enlarges on click.
        Third, they want to read, what information has been externalized, which is shown in the lower part.
        Last, we provide validation functions via voting and comment options in the bottom part of each annotation.
        Similar to established designs from social media platforms, we also instantly display the validation feedback and recorded comments.
    \subsection{Expert Feedback}
    \label{042_expert_feedback}
        In this subsection, we briefly report on how our domain experts applied annotations in order to analyze their data. 
        The feedback was gathered during several interviews and iterative application sessions and consolidated to transport key messages.

        At first, the automatic creation of annotations during data preprocessing allowed domain experts to see open issues with redundant data sources at first sight. 
        Encoding the automatically created annotations differently, depending on the existence of discrepancies and redundancies, allowed experts to focus on open issues, e.g., discrepancies.
        This reduced the overall effort to define usable data values.
        Additionally, the experts learned about the reliability of information from their clinical systems, which they considered as important for their every day work in the clinic.

        Annotations created during data cleansing, eased the comprehensibility of the editing process for the experts, as reasons for editing where given and permanently recorded.
        An analysis of the editing annotations helped to understand existing issues in the clinical process. 
        So, for example, visual acuity values are not always automatically transferred to the doctoral letter, but sometimes need to be copied by hand, increasing the risk of copying errors.

        Finally, experts noted that the ability to create notes during exploration, leads to a common information base for the discussion on the exploration results.
        On top of that experts appreciated the commenting and separation functions, that help them to confirm identified findings.

\section{Conclusion}
\label{050_discussion}
    We designed and implemented an approach to add annotations for different steps in the analysis.
    As every step has different goals, we design different annotations for each of them.
    During the design and application of the exploration tool together with the experts, we realized, that adding various annotations to the system is a large improvement of the overall analysis.
    This is particularly due to the fact, that experts can validate automatic preprocessing functions, comment on data cleansing operations, and externalize and validate their findings.
    To improve our annotations in the future, we would like to examine follow up steps in the analysis, such as verification and knowledge generation.

\printbibliography

\end{document}